\documentstyle[12 pt]{article}
\begin{document}

\bigskip
\begin{center}
{ \Large {\bf Quantum mechanics without spacetime V }}\\
{\large {\bf - Why a quantum theory of gravity should be non-linear - }}

\bigskip

\bigskip

{{\large
{\bf T. P. Singh\footnote{e-mail address: tpsingh@tifr.res.in}, Sashideep Gutti\footnote{e-mail address: sashideep@tifr.res.in} and Rakesh Tibrewala\footnote{e-mail address: rtibs@tifr.res.in} }}}

\medskip

{\it Tata Institute of Fundamental Research,}\\
{\it Homi Bhabha Road, Mumbai 400 005, India.}
\vskip 0.5cm
\end{center}

\bigskip

\begin{abstract}

\noindent If there exists a formulation of quantum mechanics which does not
refer to a background classical spacetime manifold, it then follows as a
consequence, (upon making one plausible assumption), that a quantum 
description of gravity should be necessarily non-linear. This is true 
independent of the mathematical structure used for describing such a 
formulation of quantum mechanics. A specific model which exhibits this 
non-linearity is constructed, using the language of noncommutative geometry. 
We derive a non-linear Schrodinger equation for the quantum dynamics of a 
particle; this equation reduces to the standard linear Schrodinger equation 
when the mass of the particle is much smaller than Planck mass.
It turns out that the non-linear equation found by us is 
very similar to a non-linear Schrodinger equation found by Doebner and Goldin 
in 1992 from considerations of unitary representaions of the 
infinite-dimensional group of diffeomorphisms in three spatial dimensions. 
Our analysis suggests that the diffusion constant introduced by Doebner and 
Goldin depends on the mass of the particle, and that this constant tends to 
zero in the limit in which the particle mass is much smaller than Planck 
mass, so that in this limit the non-linear theory reduces to standard linear 
quantum mechanics. A similar effective non-linear Schrodinger equation was
also found for the quantum dynamics of a system of D0-branes, by Mavromatos and Szabo.

\end{abstract}

\section{Introduction}

In a Universe in which there are no {\it classical} matter fields, it is 
desirable to have a formulation of quantum mechanics which makes no reference 
to a classical spacetime manifold. This is because the spacetime metric 
generated by the existing quantum matter fields will exhibit quantum 
fluctuations, and in accordance with the Einstein hole argument \cite{Eins}, 
such metric fluctuations will destroy the underlying classical spacetime 
manifold structure. (For a related discussion see also \cite{qm1}).

Since one could in principle have a Universe in which there are no classical
matter fields (in particular such a situation is likely to have arisen
immediately after the Big Bang), one should hence look for a way of describing
quantum mechanics without a spacetime manifold. Furthermore, such a 
description should reduce to standard quantum mechanics as and when classical
matter fields and hence a classical spacetime manifold are present.
However, in the absence of a classical spacetime manifold, one might expect interesting new physics, as the conventional  quantum mechanical description (which is true in the presence of a classical spacetime manifold) might not be valid.
In this paper we show that if such a formulation exists, it implies that a
quantum description of gravity should be non-linear. Also, it will be 
argued that such a formulation of quantum mechanics is a limiting case of a
non-linear quantum mechanics. This argument relies on one plausible
assumption that, like in classical general relativity, in a quantum 
theory of gravity also, gravity is a source for itself.

The above argument, which is straightforward and brief, will be described 
in Section 2. In Section 3, we will present the outline of a mathematical
model for such a non-linear quantum mechanics, building on previous
preliminary work which uses the language of noncommutative geometry to
construct a description of quantum mechanics without a spacetime manifold.
We derive a non-linear Schrodinger equation for a particle of mass $m$, which 
reduces to the standard linear Schrodinger equation when the mass of the 
particle is much smaller than Planck mass. It turns out that the non-linear 
equation found by us is very similar to the Doebner-Goldin equation \cite{dg}
.
\section{Why a quantum theory of gravity should be non-linear?}

Consider a `box' of elementary particles of masses $m_i$, and let each of
the masses be much smaller than the Planck mass 
$m_{Pl}\equiv (\hbar c/G)^{1/2} \sim 10^{-5}\ gms$ (this could
be a collection of electrons for instance). In this case, the dynamics
of each of the particles will follow the rules of quantum mechanics. Let 
this box be the whole Universe,
so that there will be, in the box, no classical spacetime manifold, nor a 
classical spacetime metric. Also, since $m_{i}\ll m_{Pl}$, its the same 
situation as sending $m_{Pl}$ to infinity, or letting $G\rightarrow 0$, which
means we can neglect the gravity inside the box.

The quantum dynamics of the above system will have to be described without
reference to a classical spacetime manifold. Since gravity is negligible in
this situation, whatever replaces the concept of the classical Minkowski  
manifold here can be called the `quantum version of Minkowski spacetime'.  
This quantum description of the particles in the box should become equivalent 
to standard quantum mechanics as and when a classical spacetime manifold 
becomes externally available. A classical spacetime would become externally
available if outside the box there are classical matter fields which dominate the Universe.

Consider next the case in which, to a first order approximation, the masses
$m_i$ in the box are no longer negligible, compared to Planck mass. We need 
to now take into account the `quantum gravitational field' produced by these
masses, and denoted by a set of variables, say $\eta$. Let us  
 assume that we can  associate with
the system a physical state $\Psi(\eta,m_{i})$ 
 and operators which act on the physical state. It is plausible that to this 
order of approximation the physical state is determined by a linear equation
\begin{equation}
\hat{O} \Psi(\eta,m_{i})=0
\label{lingrav}
\end{equation}
where the operator $\hat{O}$, defined on the background quantum Minkowski
spacetime, depends on the gravitational field variables $\eta$ only via the
linearized departure of $\eta$ from its `quantum Minkowski limit' and 
furthermore does not have any dependence on the physical state $\Psi$.

However, when the masses start becoming comparable to Planck mass, the
quantum gravitational field described by the state $\Psi(\eta,m_{i})$
will contribute to its own dynamics, and the operator $\hat{O}$ will pick up
non-linear corrections which depend on $\Psi(\eta,m_{i})$. In this sense, the
quantum description of the gravitational field should be via a non-linear
theory. 

One should contrast this situation with that for a quantized non-abelian
gauge theory. In the case of gravity, the operator $\hat{O}$ captures
information about the `evolution' of the quantized gravitational field, and
hence should depend on $\Psi(\eta,m_{i})$, because the gravitational field also
plays the role of describing spacetime structure. In the case of a non-abelian
gauge theory, the analog of the operator $\hat{O}$ again describes evolution
of the quantized gauge field, but the wave-functional $\Psi_{A}(A_i)$ 
describing the gauge field will not contribute to $\hat{O}$, because the
gauge-field does not describe spacetime structure. $\Psi_{A}(A_i)$ can, on
the other hand, be thought of as contributing non-linearly to a description
of the quantized geometry of the internal space on which the gauge field lives.

In approaches to quantum gravity wherein one quantizes a classical theory of
gravitation, using the standard rules of quantum theory, linearity is inherent,
by construction. Such a treatment could by itself yield a self-consistent
theory of quantum gravity. However, by requiring that there be a formulation
of quantum mechanics which does not refer to a spacetime manifold, one is led
to conclude that quantum gravity should be a non-linear theory. This happens
because we have introduced the notion of a quantum Minkowski spacetime (unavoidable when there is no classical spacetime manifold available); iterative corrections to this quantum Minkowski spacetime because of gravity bring about the non-linear dependence on the physical quantum gravitational state.

Consider now the dynamical equation which describes the motion of a 
particle $m_1$, in the absence of a classical spacetime manifold. We
assume that there are also present other particles, and in general $m_1$ 
and all the other particles together determine the `quantum gravitational
field' of this `quantum spacetime'.

The simplest situation is the one in which the masses of all the particles are
much smaller than Planck mass, so that the spacetime is a quantum version of
Minkowski spacetime, and gravity is negligible. In this case, the equation of
motion of $m_1$ will be the background independent analog of the
flat spacetime Klein-Gordon equation .
 
Keeping the mass $m_1$ small, increase the mass of the other particles so
that $m_1$ becomes a test particle, while the `quantum gravitational field'
of the other particles obeys the linear equation (\ref{lingrav}).
Here again, the equation of motion of $m_1$ will be the  
background independent analog of the curved spacetime Klein-Gordon equation,
where the `quantum gravitational field' depends linearly on the source.

Still keeping the mass $m_1$ a test particle, 
increase the mass of the other particles further, to the Planck mass range,
so that the `quantum gravitational field'
of the other particles now obeys a non-linear equation, where
the quantum gravitational field depends non-linearly on the quantum state
of these particles. In this case, the equation of motion of $m_1$ will be
 similar to the previous case but the `quantum gravitational field' now 
depends non-linearly on the quantum state of the source, the source being 
all the particles other than $m_1$.

Now increase the mass $m_1$ also, to the Planck mass range, so that its no
longer a test particle. The equation of motion of $m_1$ thus now depends
non-linearly on the quantum gravitational state of the system, where system
now includes the mass $m_1$ as well. A special case would be one where we 
remove all particles, except $m_1$, from the system. The equation of motion for
$m_1$ would then depend non-linearly on its own quantum state. This is what gives rise to non-linear quantum mechanics, which when seen from a classical spacetime manifold (as and when the latter becomes available) would be a non-linear Schrodinger equation, in the non-relativistic limit of the Klein-Gordon equation. In the next section we construct a model for such a non-linear equation, based on the language of non-commutative geometry.

\section{A model based on noncommutative geometry}

In order to build a model of quantum mechanics which does not refer to a
classical spacetime manifold, we propose to use the language of 
noncommutative differential geometry. While one could not give a foolproof 
reason that using noncommutative geometry is {\it the} correct approach,
there are plausible arguments \cite{qm1}, \cite {qm2}, \cite{hj} which,
in our opinion, make this approach an attractive one. The most attractive
reason is that in a noncommutative geometry there is a natural generalisation
of diffeomorphism invariance. 

Nonetheless, it is fair to say that, from the viewpoint of physics, 
noncommutative differential geometry is  a subject still under development,
especially with regard to Lorentzian (as opposed to Euclidean) spaces. Thus
we have here introduced an asymmetric metric - the relation of this metric  
with the more common treatment of the distance function (see for 
instance \cite{mar}) in noncommutative geometry remains to be understood. Our 
basic outlook here is that,  as a consequence of the assumptions we make, we 
derive a mass-dependent non-linear Schrodinger equation, which is in 
principle falsifiable through laboratory tests of quantum mechanics for 
mesoscopic systems; tests which could possibly be performed in the 
foreseeable future.

We shall assume that in nature, at the fundamental level, spacetime 
coordinates are non-commuting, and that one can choose to work with one of 
many possible equivalent coordinate systems. A coarse grained description of 
these noncommuting coordinates yields the familiar commuting coordinates
of the classical spacetime manifold.

 In the language of noncommutative 
geometry, diffeomorphisms of ordinary coordinates are replaced by automorphisms
of noncommuting coordinates, when one goes from a commutative geometry to a 
noncommutative one. Diffeomorphism invariance could then in principle be 
replaced by automorphism invariance if one is trying to construct a theory
of gravitation using noncommuting coordinates. Our proposal is that such
a noncommutative theory of gravity also describes quantum mechanics
\cite{hj}.

\bigskip

\centerline{\bf A possible description of quantum Minkowski spacetime}

\bigskip

A special case, which is much simpler and more tractable, as compared to the
general non-linear case, is a description of the `quantum Minkowski 
spacetime' (defined
in the previous section) using noncommuting coordinates. We recall this
construction here \cite{hj}, before going on to use this construction to
propose a model of non-linear quantum mechanics. As mentioned above, a quantum
Minkowski spacetime 
should be invoked if the masses of all the particles in the
box are small compared to Planck mass. One will have a classical spacetime 
when the masses involved become large compared to Planck mass.
For simplicity we here assume that the box has only one particle. It is not 
difficult to generalize from here to the case of many particles.

Let us start by recalling the Klein-Gordon equation 
(which we here think of as a relativistic Schrodinger equation) 
\cite{schiff} for the quantum mechanics of a particle in 2-d spacetime
\begin{equation}-\hbar^{2}
\left({\partial^{2}\over\partial t^{2}}-{\partial^{2}\over\partial x^{2}}\right)\psi=m^{2}\psi.\label{kg}
\end{equation}
This equation, after the substitution $\psi=e^{iS/\hbar}$, becomes
\begin{equation}
\left({\partial{S}\over \partial t}\right)^{2}-\left({\partial{S}\over \partial x}\right)^{2}
-i\hbar\left({\partial^{2}S\over\partial t^{2}}-{\partial^{2}S\over\partial x^{2}}\right)=m^{2}\label{hjc2}
\end{equation}
and can further be written as
\begin{equation}
p^{\mu}p_{\mu} + i\hbar {\partial p^{\mu}\over \partial x^{\mu}} = m^{2}
\label{hjp}
\end{equation}
where we have defined
\begin{equation}
p^{t}=-{\partial S\over \partial t}, \qquad p^{x} = 
{\partial S \over \partial x}
\label{pmoo}
\end{equation}
and the index $\mu$ takes the values $1$ and $2$. We assume  that $p$ gives
a definition of a `generalised momentum', in terms of the complex action $S$,
and carrying the same information as the standard momentum operator. 

Equation (\ref{hjc2}) could be thought of as a generalisation of the
classical Hamilton-Jacobi equation to the quantum mechanical case [also
for reasons which will become apparent as we proceed], where the `action' 
function $S(x,t)$ is now complex. Evidently, in (\ref{hjp}) the $\hbar$
dependent terms appear as corrections to the classical term 
$p^{\mu}p_{\mu}$. We chose to consider the relativistic case, as
opposed to the non-relativistic one, because the available spacetime
symmetry makes the analysis more transparent. Subsequently, we will
consider the non-relativistic limit.

Taking clue from the form of Eqn. (\ref{hjp}) we now suggest a model for the
dynamics of the `quantum Minkowski spacetime', in the language of two noncommuting coordinates 
$\hat{x}$ and $\hat{t}$. We ascribe to the particle a complex
`generalised momentum' $\hat{p}$,
having two components $\hat{p}^{t}$ and $\hat{p}^{x}$, which do not
commute with each other. The noncommutativity of these momentum components
is assumed to be a consequence of the noncommutativity of the coordinates,
as the momenta are defined to be the partial derivatives of the complex action
$S(\hat{x},\hat{t})$, with respect to the corresponding noncommuting 
coordinates.  

We further assume that
the coordinates $\hat{x}$ and $\hat{t}$ describe the non-commutative version
of Minkowski space, and that the noncommutative flat metric is
\begin{eqnarray}
\label{ncfm}  
\hat{\eta}_{\mu\nu} = \left(\begin{array}{cc}
                      1 & 1 \\
                      -1 & -1 \end{array} \right)
\end{eqnarray}

We now propose that the dynamics in the quantum Minkowski spacetime is given 
by the equation
\begin{equation}
\hat{p}^{\mu}\hat{p}_{\mu} = m^{2}
\label{nchj}
\end{equation}
where $\hat{p}^{\mu}$ as mentioned above are  non commuting.
Here, $\hat{p}_{\mu}=\hat{\eta}_{\mu\nu}\hat{p}^{\nu}$ is well-defined. Written
explicitly, this equation becomes
\begin{equation}
(\hat{p}^{t})^{2}-(\hat{p}^{x})^{2} + 
\hat{p}^{t}\hat{p}^{x} - \hat{p}^{x}\hat{p}^{t}  = m^{2}
\label{nce}
\end{equation} 
Eqn. (\ref{nchj}) appears an interesting and plausible proposal for the 
dynamics, because it generalizes the corresponding special relativistic 
equation to the noncommutative case. The noncommutative Hamilton-Jacobi
equation is constructed from (\ref{nce}) by defining the momentum as gradient
of the complex action function. Arguments in favour of introducing such
particle dynamics were given in \cite{hj} (Section 3 therein), motivated
by considerations of a possible spacetime symmetry that the quantum
Minkowski spacetime could have.
 
Since we know from the above arguments that the quantum Minkowski metric can be replaced by the usual Minkowski metric as and when a classical spacetime becomes available, we propose the following  rule for the 
transformation of the expression on the left of (\ref{nce}),
 whenever a classical spacetime is available: 
\begin{equation}
(\hat{p}^{t})^{2}-(\hat{p}^{x})^{2} + 
\hat{p}^{t}\hat{p}^{x} - \hat{p}^{x}\hat{p}^{t}  = ({p}^{t})^{2}-({p}^{x})^{2}
 + i\hbar {\partial p^{\mu}\over \partial x^{\mu}}
\label{nceq}
\end{equation}
Here, $p$ is the `generalised momentum' of the particle as seen from the 
commuting
coordinate system, and is related to the complex action by Eqn. (\ref{pmoo}).
This equality should be seen as two ways of  describing the same 
physics - one written in the
noncommuting coordinate system, and the other written in the standard
commuting coordinate system.

The idea here is that by using the Minkowski metric of ordinary spacetime 
one does not correctly measure the length of the `momentum' vector, because 
the noncommuting off-diagonal part is missed out. The last, 
$\hbar$ dependent term in (\ref{nceq}) provides the correction.- the origin
of this term's relation to the commutator $\hat{p}^{t}\hat{p}^{x} - \hat{p}^{x}\hat{p}^{t}$ can be understood as follows. 

Let us write this commutator by scaling all momenta with respect to Planck
momenta: define $\hat{P}^{\mu}=\hat{p}^{\mu}/P_{pl}$. Also, all lengths are scaled
with respect to Planck length: $\hat{X}^{\mu}=\hat{x}^{\mu}/L_{pl}$. Let the
components of $X^{\mu}$ be denoted as $(\hat{T}, \hat{X})$. The
commutator $\hat{P}^{T}\hat{P}^{X} - \hat{P}^{X}\hat{P}^{T}$ represents the `non-closing' of the basic quadrilateral when one
compares (i) the operator obtained by moving first along $\hat{X}$ and then along $\hat{T}$, with (ii) the operator obtained by moving first along $\hat{T}$ and then along $\hat{X}$. When seen from a commuting coordinate system, this
deficit (i.e. non-closing) can be interpreted as a result of moving to a
neighbouring point, and in the infinitesimal limit the deficit will be the
sum of the momentum gradients in the various directions. The deficit is thus given by
$i{\partial P^{\mu}/ \partial X^{\mu}}=i(L_{Pl}/P_{Pl}) {\partial p^{\mu}/ \partial x^{\mu}}$. This gives that $\hat{p}^{t}\hat{p}^{x} - \hat{p}^{x}
\hat{p}^{t}=  i\hbar {\partial p^{\mu}/ \partial x^{\mu}}$.

Hence, since the relation (\ref{nceq}) holds, 
there is equivalence between the background independent description 
(\ref{nchj}) and standard quantum dynamics given by (\ref{hjp}).

The proposal proceeds along analogous lines for four-dimensional spacetime.
The metric $\hat{\eta}_{\mu\nu}$ is defined by adding an antisymmetric part 
(all entries of which are $1$ and $-1$) to the Minkowski metric, and the 
off-diagonal contribution on the left-hand side of (\ref{nceq}) is to be set 
equal to $i\hbar {\partial p^{\mu}/ \partial x^{\mu}}$ on the right hand
side.

Our proposal for the dynamics of a `quantum Minkowski spacetime', described 
via equations
(\ref{ncfm})-(\ref{nceq}), introduces new and unusual features, and one should
ask for a justification for doing so, apart from the need to have for a 
description of quantum mechanics which does not refer to a Minkowski
spacetime. Partly, the justification comes from considerations of possible
symmetries that a noncommutative Minkowski spacetime could have \cite{hj}.
Equally importantly, the validity of the features introduced above can be 
experimentally tested by verifying or ruling out the non-linear generalisation
of the quantum Minkowski spacetime constructed below.

\bigskip

\centerline{\bf A non-linear Schrodinger equation} \bigskip

In Section 1 we argued that when the particle masses become comparable to
Planck mass, the equation of motion for the particles should become non-linear.
We now present an approximate construction for such a non-linear equation,
based on the dynamics for the `quantum Minkowski spacetime' proposed above.

The starting point for our discussion will be Eqn. (\ref{nchj}) above - we 
will assume that a natural generalisation of this equation
describes quantum dynamics when the particle mass $m$ is comparable to the Planck mass $%
m_{Pl}$. In this case, we no longer expect the metric $\eta $ to have the
`flat' form given in (\ref{ncfm}) above. The symmetric (i.e. diagonal) components of the
metric are of course expected to start depending on $m$ (this is 
usual gravity, analogous to the Schwarzschild geometry)
and in general the antisymmetric (i.e. off-diagonal) components are also expected to depend on $m
$. So long as the antisymmetric components are non-zero, we can say that
quantum effects are present. As $m$ goes to infinity, the antisymmetric
component should go to zero - since in that limit we should recover
classical mechanics. In fact the antisymmetric, off-diagonal part (let us call it $\theta
_{\mu \nu} $) should already start becoming ignorable when $m$ exceeds $m_{Pl}$%
. It is interesting to note that the symmetric part should grow with $m$, 
while the
antisymmetric part should fall with increasing $m$. There probably is a deep reason
why this is so. So, in this case, instead of eqn. (\ref{nceq}), we will have the
equation 
\begin{equation}
\label{nceq2}\hat g_{tt}(\hat p^t)^2-\hat g_{xx}(\hat p^x)^2+\hat \theta
\left( \hat p^t\hat p^x-\hat p^x\hat p^t\right) =g_{tt}({p}^t)^2-g_{xx}({p}%
^x)^2+i\hbar \theta {\frac{\partial p^\mu }{\partial x^\mu }}=m^2
\end{equation}
Here, $g$ is the symmetric part of the metric, and $\theta $ is the value of
a component of $\theta _{\mu \nu} $. 

We should now make some realistic
simplifications, in order to get an equation we can possibly tackle. Here, we are
interested in the case $m\sim m_{Pl}$. The symmetric part of the metric - $g$
- should resemble the Schwarzschild metric, and assuming we are not looking
at regions close to the Schwarzschild radius (which is certainly true for
objects of such mases which we expect to encounter in the
laboratory) we can set $g$, approximately, to unity. The key quantity is $\theta =\theta (m)$%
, and we assume that $\theta$ should be retained - it carries all the new information about any possible
non-linear quantum effects. $\theta $ in principle should also depend on the
quantum state via the complex action $S$, but we can know about the explicit
dependence of $\theta $ on $m$ and $S$ only if we know the 
dynamics of the quantum gravitational field, which
at present we do not. It is like having to know the analog of the Einstein
equations for $\theta $. But can we extract some useful conclusions just by
retaining $\theta (m)$ and knowing its asymptotic behaviour? Retaining $\theta $%
, the above dynamical equation can be written in terms of the complex action 
$S$ as follows: 
\begin{equation}
\label{hjc}\left( {\frac{\partial {S}}{\partial t}}\right) ^2-\left( {\frac{%
\partial {S}}{\partial x}}\right) ^2-i\hbar \theta (m) \left( {\frac{\partial ^2S%
}{\partial t^2}}-{\frac{\partial ^2S}{\partial x^2}}\right) =m^2
\label{rhj}
\end{equation}
This is the equation we would like to work with. We know that $\theta =1$ is
quantum mechanics, and $\theta =0$ is classical mechanics. We expect $\theta 
$ to decrease from one to zero, as $m$ is increased. It is probably more
natural that $\theta $ continuously decreases from one to zero, as one goes
from quantum mechanics to classical mechanics, instead of abruptly going
from one to zero. In that case we should expect to find experimental
signatures of $\theta $ when it departs from one, and is not too close
to zero - we expect this to happen in the vicinity of the Planck mass scale. By substituting the earlier
definition $S=-i\hbar \ln \psi $ in (\ref{rhj}) we get the following non-linear equation
for the Klein-Gordon wave-function $\psi $: 
\begin{equation}
\label{kg2}-\hbar ^2\left( {\frac{\partial ^2}{\partial t^2}}-{\frac{\partial
^2}{\partial x^2}}\right) \psi +\frac{\hbar ^2}\psi \left( 1-\frac 1\theta
\right) \left( \dot \psi ^2-\psi ^{\prime }{}^2\right) =\frac{m^2}{\theta}\psi 
\end{equation}
We are interested in working out possible consequences of the non-linearity
induced by $\theta $, even though we do not know the explicit form of $%
\theta $. 

Let us go back to the equation (\ref{hjc}). As mentioned above, in general $\theta $ will also
depend on the state $S$ but for all states $\theta $ should tend to zero for large
masses, and if we are looking at large masses we may ignore the dependence
on the state, and take $\theta =\theta (m)$. Let us define an effective Planck's
constant $\hbar _{eff}=\hbar \theta (m),$ i.e. the constant runs with the
mass $m$. Next we define an effective wave-function $\psi
_{eff}=e^{iS/\hbar _{eff}}$ . It is then easy to see from (\ref{rhj}) that the effective wave
function satisfies a linear Klein-Gordon equation
\begin{equation}
\label{psieff}-\hbar _{eff}^2\left( \frac{\partial ^2\psi _{eff}}{\partial
t^2}-\frac{\partial ^2\psi _{eff}}{\partial x^2}\right) =m^2\psi_{eff}
\end{equation}
and is related to the usual wave function $\psi $ through
\begin{equation}
\label{rel}\psi _{eff}=\psi ^{1/\theta (m)}
\end{equation}
For small masses, the effective wave-function approaches the usual wave
function, since $\theta $ goes to unity.

We would now like to obtain the non-relativistic limit for this 
equation. Evidently, this limit should be
\begin{equation}
ih_{eff}\frac{\partial\psi_{eff}}{\partial t} = 
-\frac{h_{eff}^{2}}{2m}\frac{\partial^{2}\psi_{eff}}{\partial x^{2}}.
\label{psie}
\end{equation} 
By rewriting $\psi_{eff}$ in terms of $\psi$ using the above relation 
we arrive at the following non-linear Schrodinger equation  
\begin{equation}
i\hbar\frac{\partial\psi}{\partial t} = -\frac{\hbar^{2}}{2m}\frac
{\partial^{2}\psi}{\partial x^{2}} + \frac{\hbar^{2}}{2m}(1-\theta)
\left(\frac{\partial^{2}\psi}{\partial x^{2}} - [(\ln\psi)']^{2}\psi
\right).
\label{nlse}
\end{equation}
The correction terms can also be combined so that the equation reads
\begin{equation}
i\hbar\frac{\partial\psi}{\partial t} = -\frac{\hbar^{2}}{2m}\frac
{\partial^{2}\psi}{\partial x^{2}} + \frac{\hbar^{2}}{2m}(1-\theta)
\frac{\partial^{2} (\ln\psi)}{\partial x^{2}}\psi.\label{nlse2}
\end{equation}
It is reasonable to propose that if the particle is not free, a term proportional to the potential, $V(q)\psi$, can be added to the above non-linear equation.

In terms of the complex action function $S$ defined above (\ref{hjc2}) as $\psi=e^{iS/\hbar}$ this non-linear Schrodinger equation is written as
\begin{equation}
\frac{\partial S}{\partial t} = - \frac{S'^{2}}{2m} + \frac{i\hbar}{2m}\theta(m)S''\ .
\label{nrhj}
\end{equation}
This equation is to be regarded as the non-relativistic limit of Eqn. (\ref{rhj}). 

It is clearly seen that the non-linear Schrodinger equation obtained here
results from making the Planck constant mass-dependent, in the quantum 
mechanical Hamilton-Jacobi equation. Equation (\ref{nlse}) is in principle
falsifiable by laboratory tests of quantum mechanics, and its confirmation, or
otherwise, will serve as a test of the various underlying assumptions 
of the noncommutative model. Properties of this equation, and the possible
modifications it implies for quantum mechanics of mesoscopic systems, are at
present under investigation.

\bigskip

\centerline{\bf A comparison with the Doebner-Goldin equation}

\bigskip

Remarkably enough, a non-linear Schrodinger equation very similar to (\ref{nlse}) was found some years ago by Doebner and Goldin \cite{dg}.(For a recent discussion on non-linear quantum mechanics see \cite{svet} and also \cite{adler}). They inferred their equation from a study of unitary representations of an infinite-dimensional Lie algebra of vector fields $Vect(R^{3})$ and the group of diffeomorphisms $Diff(R^{3})$. These representations provide a way to classify physically distinct quantum systems. There is a one-parameter family, labelled by a real constant $D$, of mutually inequivalent one-particle representations of the Lie-algebra of probability and current densities. The usual one-particle Fock representation is the special case $D=0$.
The probability density $\rho$ and the current density ${\bf j}$ satisfy, not the continuity equation, but a Fokker-Planck equation
\begin{equation}
\frac{\partial \rho}{\partial t} = - {\bf \nabla.j} + D\nabla^{2}\rho.
\label{fp}
\end{equation}    
A linear Schrodinger equation cannot be consistent with the above Fokker-Planck equation with $D\ne 0$, but Doebner and Goldin found that the following non-linear Schrodinger equation leads to the above Fokker-Planck equation
\begin{equation}
i\hbar \frac{\partial\psi}{\partial t} = -\frac{\hbar^{2}}{2m}\nabla^{2}\psi + iD\hbar\left(
\nabla^{2}\psi + \frac{|\nabla\psi|^{2}}{|\psi|^{2}}\psi\right).
\end{equation}

The Doebner-Goldin equation should be compared with the equation (\ref{nlse}) found by us. Although we have considered a 2-d case, and although there are some important
differences, the similarity between the two equations is striking, considering
that the two approaches to this non-linear equation are, at least on the face of it, quite different. It remains to be seen as to what is the connection between the representations of $Diff({R^{3}})$, quantum mechanics,  and the antisymmetric part $\theta$ of the asymmetric metric introduced by us. (A similar effective non-linear Schrodinger equation was
also found for the quantum dynamics of a system of D0-branes, by Mavromatos and 
Szabo \cite{mav}).

The comparison between the two non-linear equations suggests the following relation between the new constants $D$ and $\theta$
\begin{equation}
D\sim \frac{\hbar}{m}(1-\theta).
\label{dtheta}
\end{equation} 
There is a significant difference of an $i$ factor in the correction terms in
the two equations, and further, the relative sign of the two correction terms 
is different in the two equations, and in the last term we do not have absolute values in the numerator and denominator, unlike in the Doebner-Goldin equation.
The origin of these differences is not clear to us, nonetheless the similarity
between the two equations is noteworthy and we believe this aspect should be
explored further. It is encouraging that there is a strong parallel between the limits $D\rightarrow 0$ and $\theta\rightarrow 1$ - both limits correspond to standard linear quantum mechanics. In their paper Doebner and Goldin note that  the constant $D$ could be different for different particle species. In the
present analysis we clearly see that $\theta(m)$ is labelled by the mass of the particle. 

If we substitute $\psi=e^{iS/\hbar}$ in the Doebner-Goldin equation we get the
following corrected quantum mechanical Hamilton-Jacobi equation (written for 
one spatial dimension), after defining $D=-i\hbar(1-\theta)/2m$
\begin{equation}
\frac{\partial S}{\partial t} = - \frac{S'}{2m}
\left[\theta S' + (1-\theta) S^{*'}\right]
 + \frac{i\hbar}{2m}\theta(m)S''\ .
\label{nrhj2}
\end{equation}
This equation should be compared with our Eqn. (\ref{nrhj}).

For comparison with Doebner and Goldin we also write down the corrections to
the
continuity equation which follow as a consequence of the non-linear terms in
(\ref{nlse}). These can be obtained by first noting, from (\ref{psie}), that
the effective wave-function $\psi_{eff}$ obeys the following continuity 
equation
\begin{equation}
\frac{\partial\ }{\partial t}\left( \psi^{*}_{eff}\psi_{eff}\right)
-\frac{i\hbar_{eff}}{2m}\left(\psi_{eff}^{*}\psi_{eff}' - 
\psi_{eff}\psi_{eff}^{*'}\right)' = 0.
\end{equation}
By substituting $\psi_{eff}=\psi^{1/\theta (m)}$ and $\hbar_{eff}=\hbar\theta(m)$ in this equation we get the following corrections to the continuity 
equation for the probability and current density constructed from $\psi(x)$ 
\begin{equation}
\frac{\partial\ }{\partial t}\left( \psi^{*}\psi\right)
-\frac{i\hbar}{2m}\left(\psi^{*}\psi' - 
\psi\psi^{*'}\right)' = \frac{\hbar(1-\theta)}{m}|\psi|^{2}\phi''
\label{cm}
\end{equation}
where $\phi$ is the phase of the wave-function $\psi$, i.e. $\phi=Re(S)/\hbar$.
It is interesting that the phase enters in a significant manner in the
correction to the continuity equation. This equation should be contrasted
with Eqn. (\ref{fp}). The fact that the evolution is not norm-preserving
when the mass becomes comparable to Planck mass perhaps suggests that the
appropriate description could be in terms of the effective wave-function
$\psi_{eff}$.

\section{Discussion}

In our model, based on noncommuting coordinates, we have introduced a few
new assumptions, a detailed understanding of which should emerge
when the relationship between noncommutative differential geometry and quantum 
gravity is better understood. Thus we have proposed the notion of a 
generalised momentum, an asymmetric metric, and a generalised energy-momentum
equation. It should be said that these ideas do find support in the 
generalisation from diffeomorphism invariance, to automorphism invariance,
when one goes over to noncommuting coordinates, from commuting ones \cite{hj}.
Nonetheless, we would like to take the stand that such underlying ideas should
be put to experimental test, by confirming or ruling out, in the laboratory,
the non-linearities predicted in the vicinity of the Planck mass scale.
Such tests could also help decide whether or not quantum gravity is a linear
theory.

There are very stringent experimental constraints on possible non-linearities 
in quantum mechanics, in the atomic domain. However, quantum mechanics has
not really been tested in the mesoscopic domain, say for an object having
a mass of a billionth of a gram. Such experiments are very difficult to
perform with current technology, but given the advances in nanotechnology,
perhaps not impossible to carry out in the foreseeable future. While most
people do not expect any surprises, we feel that if there are theoretical 
models predicting non-linearity precisely in the domain which has not
been experimentally explored thus far, we should keep an open mind about
the possible outcome. At the very least, such experiments can put strong
bounds on non-linearities in mesoscopic quantum mechanics.       

In the context of the model discussed here,
how can we possibly interpret the effective wave function $\psi_{eff}$? 
It would appear
correct to say that in the present model, the correct physics is described
by the effective wave function, and not by the usual wave function, even
though the two carry the same information. Now for large masses, $\theta $
goes to zero, this makes $1/\theta $ very large. If we consider a $\psi $
which is a superposition of two eigenstates of an observable which we are
measuring, we can see that the exponent $1/\theta $ will lead to extremely
rapid oscillations between the two eigenstates. In this case we can possibly
set the oscillatory phase to zero - this essentially destroys superpositions
between states. This seems to behave like a decohering mechanism which
destroys superpositions, replacing sum of amplitudes by sum of
classical probabilities. Furthermore, the $1/\theta $ term will lead to a
very rapid fall-off of the effective wave function squared with distance,
possibly providing an explanation for spatial localization of macroscopic
systems.

The presence of the $\theta $ term may thus provide for decoherence. What is
not clear yet, though, is the origin of the Born probability rule. A model
such as the present one should provide an explanation for the collapse of
the wave function according to the Born rule. Also, the $\theta $ term
should be able to explain why the quantum system jumps into one of the
eigenstates, upon measurement, even though originally its in a superposition
of the eigenstates. This issue, as well as the possibility of a connection with
the Doebner-Goldin equation, is under investigation. Another aspect under
study, and discussed briefly in \cite{hj}, is the nature of commutation
relations for position and momenta, in noncommuting coordinate systems.

It is a pleasure to acknowledge useful discussions with C. S. Unnikrishnan
and Cenalo Vaz, and useful comments from George Svetlichny.

\end{document}